\documentclass[aps,pra,twocolumn,10pt,longbibliography]{revtex4-2}
\usepackage[T1]{fontenc}
\usepackage{graphicx}
\usepackage{amsmath,amssymb,amsthm,mathtools}
\usepackage{physics}
\usepackage{mathpazo}
\usepackage{xspace,xcolor}
\usepackage[colorlinks=true,citecolor=blue,urlcolor=blue]{hyperref}
\newcommand{\id}{\openone}
\newcommand{\Sep}{\mathcal S}
\newcommand{\dhs}{D_{\mathrm{HS}}}
\newcommand{\dhm}{D_{\mathrm{HS}_{\min}}}
\newcommand{\dhssq}{D^2_{\mathrm{HS}}}
\newcommand{\dhssqm}{D^2_{\mathrm{HS}_{\min}}}

\newcommand{\rcss}{\rho_{\mathrm{CSS}}}

\newtheorem{theorem}{Theorem}
\newtheorem{lemma}[theorem]{Lemma}
\newtheorem{definition}{Definition}
\makeatletter
\AtBeginDocument{\immediate\write\@auxout{\string\citation{minhsdControl}}}
\makeatother

\begin{document}
\title{Minimum Hilbert--Schmidt distance for pure states of Schmidt rank two}
\author{Palash Pandya}
\email{ppandya@cft.edu.pl}
\affiliation{Center for Theoretical Physics, Polish Academy of Sciences,
Aleja Lotnik\'{o}w 32/46, 02-668 Warsaw, Poland}

\begin{abstract}
The Hilbert--Schmidt norm is not contractive under general completely
positive trace-preserving maps. This fact alone does not decide whether
distance from the separable set decreases under a restricted class of
entanglement transformations. For bipartite pure states with two nonzero
Schmidt coefficients, we give a self-contained derivation of the minimum
distance and its closest separable state. We prove global optimality of
the boundary construction previously conjectured by Witte and Trucks and
show that the result is unchanged by embedding the state in larger local
Hilbert spaces. An explicit product-state decomposition certifies
separability, and the embedded optimizer also minimizes distance over
the full PPT set. The squared distance has a simple branch above a
golden-ratio threshold and a cubic branch below it. A bound on the
physical cubic root gives a direct proof that both the distance and its
square are non-increasing under deterministic LOCC transformations
between pure states of Schmidt rank at most two. Unrestricted
two-qubit PPT optimization independently verifies the formulas, while
Rust Gilbert iterations give separable upper bounds.
\end{abstract}
\maketitle

\section{Introduction}
Entanglement can be quantified geometrically by minimizing a suitable
distance from the convex set of separable states
\cite{quantifyingEntanglement,vedral-purification-procedures}.
The Hilbert--Schmidt norm makes this geometry Euclidean on the real
vector space of Hermitian operators. For a single qubit the state space
is the Bloch ball. In higher dimensions, positivity restricts physical
generalized Bloch vectors to a proper subset of the enclosing ball.
Rank-deficient mixed states lie on the boundary of this physical region,
strictly inside the enclosing sphere.
The norm itself is easy to evaluate, although optimization over the
separable set remains a separate problem.

Witte and Trucks introduced the minimum squared Hilbert--Schmidt
distance as an entanglement quantity \cite{witte-trucks}. Their
Proposition~3 gives the pure two-qubit expression
$4\alpha^2\beta^2/3$ and its golden-ratio domain. For the remaining
states their Conjecture~4 specifies a boundary-state parametrization and
a cubic stationarity equation. Their Remark~8 also states that the
pure-state ordering agrees with that of entropy of entanglement.
These are direct precedents for the two branches studied here.
We provide a proof of global optimality of that construction, an
explicit radical expression for the physical root, and a proof that
unused local dimensions do not change the optimum. We also give an
explicit derivative inequality establishing the deterministic
pure-state LOCC (Local Operations and Classical Communication) ordering. We do not claim that the two-branch
construction first appears in this work.

Ozawa showed that Hilbert--Schmidt distance between states need not
contract under completely positive trace-preserving (CPTP) maps
\cite{ozawa}. Norm contractivity was further characterized by
P\'erez-Garc\'ia \emph{et al.}\ \cite{garcia-wolf-contractivity}; their
results distinguish unrestricted operators from traceless Hermitian
differences of states. Failure of norm contractivity removes a standard
sufficient route to an entanglement monotone, but does not by itself
settle a restricted pure-state ordering problem for the minimized
distance. It is that restricted problem which we address.

For comparison, negativity and concurrence admit explicit formulas for
two-qubit states \cite{negativity,concurrence}, and entanglement of
formation is computable through concurrence
\cite{concurrence-2-EOF}. Numerical approaches to separable-distance
problems include Gilbert-type convex optimization
\cite{otfried-guenhe-hsd,HSD-1,wiesniak2020distance} and neural-network
approximations \cite{hsd-neural-net}. The closest separable state also
determines a supporting entanglement witness
\cite{Bertlmann2001AGP,Pittenger2002GeometryOE}. Tangency alone does not
make such a witness optimal in the sense of detecting a maximal set of
entangled states \cite{witness-optimization}.

Throughout, Schmidt rank two refers to a \emph{pure} state with two
nonzero Schmidt coefficients. We include rank-one states as the
zero-entanglement endpoint. Our monotonicity statement concerns
deterministic LOCC conversions between specified pure states. We make
no mixed-state or on-average selective-LOCC monotonicity claim.

Section~\ref{sec:hsd} defines the distance. In
Sec.~\ref{sec:min-hsd-2-qubit} we derive the two branches and give the
closest state and numerical checks. Section~\ref{sec:hsd-under-LOCC}
states and proves the deterministic ordering using Nielsen's theorem
\cite{Nielsen_1999}. The symmetry, saturation and embedding arguments
are given in Appendix~\ref{app:commutator}; the cubic-root bound and
derivative inequality are proved in Appendix~\ref{app:derivative}.

\section{Minimum Hilbert--Schmidt distance}
\label{sec:hsd}
For density operators $\rho$ and $\sigma$, define
\begin{equation}
 \dhs(\rho,\sigma)=\norm{\rho-\sigma}_2
 =\sqrt{\Tr[(\rho-\sigma)^2]}.
\end{equation}
Here $\norm{X}_2=\sqrt{\Tr(X^\dagger X)}$. The general
distance-to-separability construction is
\begin{equation}
\label{eq:define-ent-dist}
 E(\rho)=\min_{\sigma\in\Sep}D(\rho,\sigma),
\end{equation}
where $\Sep$ is the set of separable density operators in the specified
bipartite Hilbert space.
\begin{definition}
The minimum Hilbert--Schmidt distance is
$\dhm(\rho)=\min_{\sigma\in\Sep}\norm{\rho-\sigma}_2$. Its square is
\begin{equation}
\label{eq:HS-measure}
 \dhssqm(\rho)=\min_{\sigma\in\Sep}\norm{\rho-\sigma}_2^2.
\end{equation}
\end{definition}
Both quantities vanish precisely for separable states. Compactness and
convexity of $\Sep$ imply that the minimizer exists and is unique,
because the squared norm is strictly convex. Local unitaries preserve
the feasible set and the distance.

For two qubits, separability is equivalent to positivity of the partial
transpose (PPT) \cite{PPT-1,PPT-2}. This permits an independent
semidefinite optimization check. In larger dimensions PPT is necessary
but not sufficient in general. Our extension to larger local dimensions
therefore uses a supporting-hyperplane proof rather than assuming
PPT sufficiency there.

\section{States with Schmidt rank two}
\label{sec:min-hsd-2-qubit}
\subsection{Two-qubit reduction}
\label{sec:schmidt-decompostion}
By local unitary invariance we may write
\begin{equation}
 \begin{gathered}
 \ket\psi=\alpha\ket{00}+\beta\ket{11},\quad \rho=\op\psi,\\
 0\leq\alpha\leq\beta,\qquad \alpha^2+\beta^2=1.
 \end{gathered}
\end{equation}
Partial transposition preserves the Hilbert--Schmidt norm. For a general
Schmidt decomposition $\sum_i\alpha_i\ket{ii}$, the nonzero
partial-transpose eigenvalue multiset is
\begin{equation}
 \{\alpha_i^2\}_i\ \cup\
 \{+\alpha_i\alpha_j,-\alpha_i\alpha_j\}_{i<j}.
\end{equation}
The corresponding normalized eigenvectors are $\ket{ii}$ and
$(\ket{ij}\pm\ket{ji})/\sqrt2$; additional ambient dimensions add zero
eigenvalues \cite{Rana-partial-transpose-EV}. In this convention
negativity is $\sum_{i<j}\alpha_i\alpha_j$, and generalized pure-state
concurrence satisfies $C^2=4\sum_{i<j}\alpha_i^2\alpha_j^2$.

The spectral redistribution method of Verstraete \emph{et al.}\
\cite{verstraete-method} first projects $\rho^\Gamma=UDU^\dagger$
onto positive trace-one operators: the eigenvalues are shifted and
clipped as $e_i=\max(d_i+\ell,0)$, with $\sum_i e_i=1$.
The inverse-transposed candidate is $(UEU^\dagger)^\Gamma$.
It is a PPT solution only if this last matrix is also positive.
The extra positivity condition is essential below.

For pure inputs, a local phase twirl, interchange of the two Schmidt
supports, and complex conjugation justify a common eigenbasis in the
partially transposed picture. For rank two a further
\emph{trace-preserving} descent proves saturation of the relevant PPT
inequality. Appendix~\ref{app:commutator} consequently reduces the full
two-qubit problem to
\begin{equation}
\label{eq:rank2-state}
 \rho_{\mathrm{red}}=
 \begin{pmatrix}
 e_1&0&0&e_3/2\\
 0&e_3/2&0&0\\
 0&0&e_3/2&0\\
 e_3/2&0&0&e_2
 \end{pmatrix},
\end{equation}
with $e_i\geq0$, $e_1+e_2+e_3=1$. Its eigenvalues are
\begin{multline}
\label{eq:ev-rhored}
 \left\{\frac{e_3}{2},\frac{e_3}{2},
 \frac{e_1+e_2-\sqrt{(e_1-e_2)^2+e_3^2}}{2},\right.\\
 \left.\frac{e_1+e_2+\sqrt{(e_1-e_2)^2+e_3^2}}{2}\right\}.
\end{multline}
Thus the remaining condition for positivity of
$\rho_{\mathrm{red}}$, after inverse partial transpose, is
$4e_1e_2-e_3^2\geq0$.

\subsection{The simple branch and its boundary}
For $\alpha\beta>0$ write
\begin{equation}
 e_1=\alpha^2-a\alpha\beta,\quad
 e_2=\beta^2-b\alpha\beta,\quad e_3=(a+b)\alpha\beta.
\end{equation}
The variables $a,b$ are constrained through the nonnegativity, trace and
PSD conditions on the $e_i$; no strict positivity of eigenvalues is
assumed. The squared distance is
\begin{equation}
\label{eq:hsd-schmidt}
 \dhssq(\rho,\rho_{\mathrm{red}})
 =[1+a^2+b^2+(1-a-b)^2]\alpha^2\beta^2.
\end{equation}
Completing squares gives
\begin{align}
 1+a^2+b^2+(1-a-b)^2={}&\frac43+(a-\tfrac13)^2
 +(b-\tfrac13)^2\nonumber\\
 &+(a+b-\tfrac23)^2.
\end{align}
The unconstrained minimum is therefore unique at $a=b=1/3$.
Put $N=\alpha\beta$. The candidate coefficients and determinant are
\begin{align}
 e_1&=\alpha^2-N/3,\quad e_2=\beta^2-N/3,
 \quad e_3=2N/3,\label{eq:simple-coefficients}\\
 4e_1e_2-e_3^2&=4N(N-1/3).\label{eq:centroid-det}
\end{align}
For $\alpha>0$, feasibility holds exactly when $N\geq1/3$, or
\begin{equation}
\label{eq:threshold}
 \alpha\geq\alpha_*:=\sqrt{\frac{3-\sqrt5}{6}}
 =\sqrt{\frac{2-\varphi}{3}},\qquad
 \varphi=\frac{1+\sqrt5}{2}.
\end{equation}
In this interval the squared minimum is $4\alpha^2\beta^2/3$.
If the full reflected interval $0\leq\alpha\leq1$ is plotted, the
determinant is negative on
\begin{equation}
\label{eq:neg-interval}
 (0,\alpha_*)\ \cup\ (\sqrt{1-\alpha_*^2},1).
\end{equation}
The simple branch and this threshold agree with Proposition~3 of
Ref.~\cite{witte-trucks}.

\begin{figure}[t]
 \includegraphics[width=\linewidth]{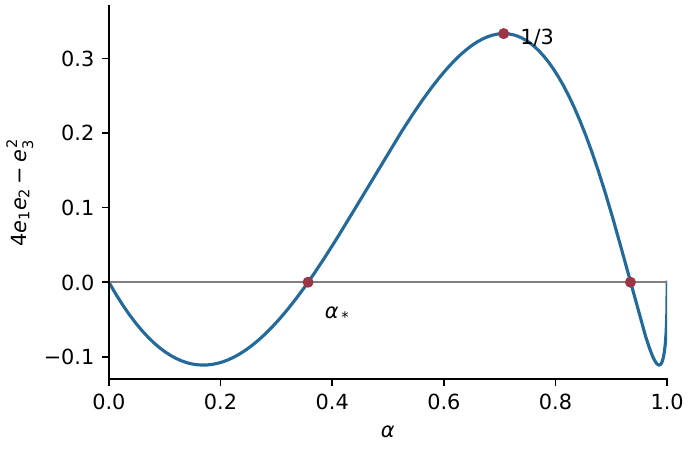}
 \caption{The determinant condition $4e_1e_2-e_3^2$ for equal
 redistribution. The peak is $1/3$ at $\alpha=1/\sqrt2$;
 the nontrivial zero crossings are $\alpha_*$ and
 $\sqrt{1-\alpha_*^2}$. The squared minimum is
 $4\alpha^2\beta^2/3$ wherever this candidate is feasible.}
 \label{fig:eigenvalue-schmidt}
\end{figure}

\subsection{The boundary branch}
When $0<\alpha<\alpha_*$, the unconstrained minimizer is infeasible.
Appendix~\ref{app:commutator} shows that the optimum has $e_i>0$ and
lies on $4e_1e_2=e_3^2$. Since $e_3\geq0$ and $\sum_i e_i=1$,
\begin{equation}
\label{eq:e1e2eq1}
 \sqrt{e_1}+\sqrt{e_2}=1.
\end{equation}
It is convenient to set
\begin{equation}
\label{eq:boundary-param}
 t=\sqrt{e_1},\quad e_1=t^2,\quad e_2=(1-t)^2,
 \quad e_3=2t(1-t).
\end{equation}
The boundary objective is
\begin{multline}
\label{eq:boundary-objective}
 F(\alpha,t)=\alpha^2\beta^2+(\alpha^2-t^2)^2
 +[\beta^2-(1-t)^2]^2\\
 +[\alpha\beta-2t(1-t)]^2.
\end{multline}
The physical stationary root solves
\begin{equation}
\label{eq:boundary-cubic}
 6t^3-9t^2+(4+2\alpha\beta)t-\alpha^2-\alpha\beta=0.
\end{equation}
Appendix~\ref{app:derivative} proves that there is exactly one root in
$0<t<1/2$, that this root is the global boundary minimizer, and that
$t\leq\alpha\beta/2$. Equation~\eqref{eq:boundary-cubic} is identical
to Eq.~(6) of Ref.~\cite{witte-trucks} after normalization of the Schmidt
coefficients. The global reduction here establishes optimality over all
two-qubit PPT, and hence separable, states.

For completeness, the root can be given in radicals. With
\begin{equation}
 g_{\alpha\beta}=9(\beta^2-\alpha^2)
 +2\sqrt{16\alpha^3\beta^3-93\alpha^2\beta^2+3\alpha\beta+20},
\end{equation}
the corresponding coefficient is
\begin{equation}
\label{eq:e2min}
 e_2=\frac1{36}\left(3+
 \frac{1-4\alpha\beta}{\sqrt[3]{g_{\alpha\beta}}}
 +\sqrt[3]{g_{\alpha\beta}}\right)^2,
\end{equation}
where cube roots are real. The auxiliary redistribution variable is
\begin{equation}
\label{eq:bmin}
 b_{\min}=\frac{\beta^2-e_2}{\alpha\beta}
 \qquad(0<\alpha\leq\alpha_*).
\end{equation}
At $\alpha=0$ use the product-state coefficients $(e_1,e_2,e_3)=(0,1,0)$
directly, rather than the auxiliary expression with its zero denominator.

\subsection{Minimum distance and closest separable state}
For $0\leq\alpha\leq1/\sqrt2$ and $\beta=\sqrt{1-\alpha^2}$,
\begin{equation}
\label{eq:min-hsd-def}
 \dhssqm(\alpha,\beta)=
 \begin{cases}
 0,&\alpha=0,\\
 D(\alpha,\beta),&0<\alpha<\alpha_*,\\
 \frac43\alpha^2\beta^2,&\alpha_*\leq\alpha\leq1/\sqrt2,
 \end{cases}
\end{equation}
where $D$ denotes the \emph{squared-distance} boundary branch,
\begin{multline}
\label{eq:boundary-distance}
 D(\alpha,\beta)=\alpha^2\beta^2+
 [\alpha^2-(1-\sqrt{e_2})^2]^2\\
 +(\beta^2-e_2)^2+(\alpha\beta-2\sqrt{e_2}+2e_2)^2,
\end{multline}
and $e_2$ is given by Eq.~\eqref{eq:e2min}. At the junction both
branches give $4/27$, and the unsquared distance is
$\dhm=\sqrt{\dhssqm}$.

The closest separable state is Eq.~\eqref{eq:rank2-state}. Writing
$\boldsymbol e=(e_1,e_2,e_3)$, its \emph{piecewise} coefficients are
\begin{equation}
\label{eq:css-piecewise}
 \boldsymbol e=
 \begin{cases}
 (0,1,0),&\alpha=0,\\
 (t^2,(1-t)^2,2t(1-t)),&0<\alpha<\alpha_*,\\
 (\alpha^2-N/3,\beta^2-N/3,2N/3),&\alpha\geq\alpha_*.
 \end{cases}
\end{equation}
Equivalently, define $\ket{\eta_1}=\ket{00}$,
$\ket{\eta_2}=\ket{11}$ and
$\ket{\eta_{3,4}}=(\ket{01}\pm\ket{10})/\sqrt2$. Then
\begin{equation}
 \rho_{\mathrm{red}}^\Gamma=\sum_{m=1}^3e_m\op{\eta_m},
 \qquad \rcss=(\rho_{\mathrm{red}}^\Gamma)^\Gamma.
\end{equation}
The boundary coefficients must not be used above $\alpha_*$.
For example, the Bell state has $(e_1,e_2,e_3)=(1/3,1/3,1/3)$
and $\dhssqm=1/3$.

For two-qubit states the PPT and separable sets coincide
\cite{PPT-2}. The optimizer above is therefore simultaneously the
closest PPT state and the closest separable state; separability itself
is a consequence of this established criterion. The following explicit
decomposition gives a constructive verification in both branches.

\subsubsection{Explicit separable decomposition}
\label{sec:explicit-decomposition}
For $\alpha>0$, the coefficients in Eq.~\eqref{eq:css-piecewise}
satisfy $e_1e_2>0$. Define
\begin{equation}
\label{eq:decomp-weights}
 s=\frac{e_3}{2\sqrt{e_1e_2}},\qquad
 p=s(\sqrt{e_1}+\sqrt{e_2})^2,
\end{equation}
and normalized local vectors
\begin{equation}
\begin{gathered}
 |\phi_\theta\rangle=u|0\rangle+v e^{i\theta}|1\rangle,\\
 u^2=\frac{\sqrt{e_1}}{\sqrt{e_1}+\sqrt{e_2}},\qquad
 v^2=\frac{\sqrt{e_2}}{\sqrt{e_1}+\sqrt{e_2}},
\end{gathered}
\end{equation}
where $u,v\geq0$. Writing $\theta_k=2\pi k/3$ and
$|\Phi_k\rangle=|\phi_{\theta_k}\rangle\otimes
|\phi_{-\theta_k}\rangle$, we obtain
\begin{equation}
\label{eq:product-decomposition}
\begin{aligned}
 \rcss={}&\frac p3\sum_{k=0}^2|\Phi_k\rangle\langle\Phi_k|\\
 &+(1-s)e_1|00\rangle\langle00|
 +(1-s)e_2|11\rangle\langle11|.
\end{aligned}
\end{equation}
Indeed, positivity requires $e_3^2\leq4e_1e_2$, so $0\leq s\leq1$.
The weights are nonnegative and sum to one because
\begin{equation}
 p+(1-s)(e_1+e_2)=e_1+e_2+e_3=1.
\end{equation}
The three-phase average cancels terms proportional to
$e^{\pm i\theta_k}$ and $e^{\pm2i\theta_k}$. Its remaining entries
follow from $pu^4=se_1$, $pv^4=se_2$ and $pu^2v^2=e_3/2$,
which reproduce Eq.~\eqref{eq:rank2-state}.
Thus Eq.~\eqref{eq:product-decomposition} is a mixture of at most
five pure product states. On the boundary branch,
$e_3=2\sqrt{e_1e_2}$ and $\sqrt{e_1}+\sqrt{e_2}=1$, so
$s=p=1$ and only the three phase-related terms remain.
At $\alpha=0$, $\rcss=|11\rangle\langle11|$ directly.
This phase-average construction is a special case of the established
separability framework for conjugate local diagonal unitary invariant
states \cite{JohnstonMacLean2019}.

For larger local dimensions, apply the Schmidt-basis local isometries
to this same decomposition. Appendix~\ref{app:embedding} proves that
the embedded state remains the global closest state over the full PPT
set, including candidates outside the Schmidt support. It follows that
for every bipartite pure state of Schmidt rank at most two,
\begin{equation}
\label{eq:ppt-sep-equality}
 \min_{\sigma\in\mathrm{PPT}}\norm{\rho-\sigma}_2^2
 =\min_{\sigma\in\Sep}\norm{\rho-\sigma}_2^2,
\end{equation}
with the same unique, separable optimizer. This statement concerns
Schmidt rank at most two and does not assert equality for arbitrary
higher-rank pure inputs.

\subsection{Relation to concurrence and negativity}
\label{sec:concurrence-negativity}
For the pure states considered here, concurrence and negativity are
\begin{equation}
\label{eq:concurrence-negativity}
 C=2\alpha\beta=2N,\qquad 0\leq C\leq1,
\end{equation}
where $N=(\norm{\rho^\Gamma}_1-1)/2$ is the unnormalized negativity
\cite{negativity,concurrence-2-EOF}. Thus either quantity determines
the ordered Schmidt spectrum:
\begin{equation}
 \alpha^2=\frac{1-\sqrt{1-C^2}}2,\qquad
 \beta^2=\frac{1+\sqrt{1-C^2}}2.
\end{equation}
The minimum distance can consequently be expressed as a function of
$C$ alone, or equivalently of $N$ alone. In particular,
\begin{equation}
\label{eq:hsd-concurrence}
 \dhssqm=
 \begin{cases}
 H(C),&0\leq C<2/3,\\
 C^2/3,&2/3\leq C\leq1.
 \end{cases}
\end{equation}
On the simple branch this gives the proportionality
\begin{equation}
\label{eq:hsd-simple-concurrence}
 \dhm=\frac C{\sqrt3}=\frac{2N}{\sqrt3}
 \qquad(N\geq1/3).
\end{equation}

For the boundary branch, put $\delta=\sqrt{1-C^2}$ and
$y=1-2t$, with $t$ as in Eq.~\eqref{eq:boundary-param}.
The stationary equation becomes
\begin{equation}
\label{eq:concurrence-cubic}
 3y^3+(2C-1)y-2\delta=0.
\end{equation}
Substituting $t=(1-y)/2$ into the boundary objective gives
\begin{equation}
 F=\frac78-\frac C2-\delta y
 +\frac{2C-1}{4}y^2+\frac38y^4.
\end{equation}
Eliminating the quartic term with
Eq.~\eqref{eq:concurrence-cubic} yields the compact value
\begin{equation}
\label{eq:hsd-concurrence-boundary}
 H(C)=\frac{7-4C-6\delta y+(2C-1)y^2}{8}.
\end{equation}
There is no further minimization: the physical root is
\begin{equation}
\label{eq:concurrence-root}
\begin{gathered}
 G(C)=9\sqrt{1-C^2}+\sqrt{8C^3-93C^2+6C+80},\\
 y=\frac13\left(\sqrt[3]{G(C)}
       +\frac{1-2C}{\sqrt[3]{G(C)}}\right),
\end{gathered}
\end{equation}
where cube roots are real. The formula includes $H(0)=0$ by
continuity and gives $H(2/3)=4/27$, matching the simple branch.
Replacing $C$ by $2N$ gives the negativity-only expression in both
regimes. Near the product endpoint,
\begin{equation}
\label{eq:hsd-concurrence-expansion}
 \dhssqm=\frac38C^2-\frac{15}{128}C^3+O(C^4).
\end{equation}
Hence proportionality to concurrence or negativity holds only on
the simple branch. The ordering consequence is established in
Sec.~\ref{sec:hsd-under-LOCC}.

\subsection{Supporting entanglement witness}
Let $\tau=\rcss$ and assume $\alpha>0$. The projection inequality is
\begin{equation}
 \Tr[(\tau-\rho)(\sigma-\tau)]\geq0
 \qquad\text{for every }\sigma\in\Sep.
\end{equation}
Consequently,
\begin{equation}
\label{eq:supporting-witness}
 \mathcal W=\frac{\tau-\rho-c_0\id}{\norm{\tau-\rho}_2},
 \qquad c_0=\Tr[(\tau-\rho)\tau],
\end{equation}
is nonnegative on separable states, tangent at $\tau$, and negative on
$\rho$: its unnormalized numerator has expectation $-\dhssqm(\rho)$
on the input. This is a supporting witness
\cite{Bertlmann2001AGP,Pittenger2002GeometryOE}; we do not infer
detection-set optimality merely from tangency
\cite{witness-optimization}. At the product endpoint the denominator
vanishes and no entanglement witness is required.

\subsection{Numerical validation}
\label{sec:numerics}
We check the analytical result against the unrestricted two-qubit
PPT optimization
\begin{equation}
 \begin{gathered}
 \min_X\norm{\rho-X}_2^2\quad\text{subject to}\\
 X\succeq0,\quad X^\Gamma\succeq0,\quad\Tr X=1.
 \end{gathered}
\end{equation}
We solve
this problem using CVXPY and CLARABEL for 75 inputs spanning
$0\leq\alpha\leq1/\sqrt2$, including the branch junction and nearby
points. The largest discrepancy from the analytical \emph{squared}
distance is $3.78\times10^{-12}$. Five additional $3\times3$ PPT
optimizations of embedded rank-two inputs agree to within
$5.91\times10^{-12}$. We also independently check the cubic-root
expression, derivative, product-state decomposition and concurrence
representation.

\begin{figure}[t]
 \includegraphics[width=\linewidth]{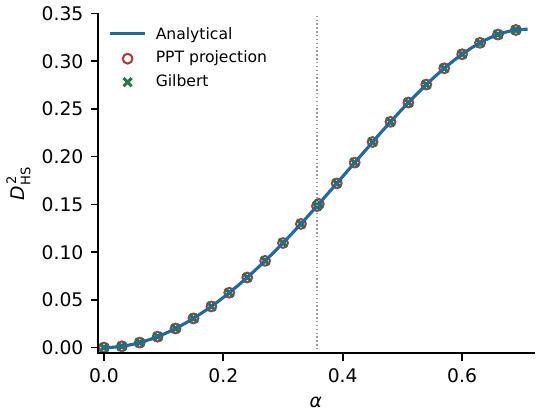}
 \caption{Squared Hilbert--Schmidt distance from the analytical formula,
 unrestricted two-qubit PPT optimization and Gilbert's algorithm on
 the same 75-input grid. Every third numerical sample is shown.
 The dotted line marks $\alpha_*$.}
 \label{fig:schmidt-comparison}
\end{figure}

A second comparison uses Gilbert's algorithm to construct separable
approximations \cite{HSD-1,wiesniak2020distance,gilbert-opt-rs}.
Each iterate is a convex mixture of product states and therefore
provides an upper bound on the minimum distance. Figure~\ref{fig:schmidt-comparison}
shows the analytical, PPT and Gilbert results on the same 75-input
grid. Each run stops upon reaching $10^5$ accepted corrections,
$10^8$ trials, or a squared-distance gap of $10^{-8}$ from the
analytical optimum, whichever occurs first. The analytical reference
is used only for stopping, not for trial generation or iterate updates.
The largest squared-distance excess above the analytical squared
distance is $8.40\times10^{-5}$.
Finite Gilbert results remain upper bounds and do not replace the
optimality proof.

The scripts, detailed numerical settings and complete run data are
available in the accompanying code repository \cite{minhsd-code}.

\section{Deterministic pure-state LOCC monotonicity}
\label{sec:hsd-under-LOCC}
Nielsen's theorem states that a deterministic LOCC transformation
$\ket\psi\longrightarrow\ket\phi$ between bipartite pure states is
possible if and only if $\lambda_\psi\prec\lambda_\phi$
\cite{Nielsen_1999}. Here the $\lambda$ vectors contain the squared
Schmidt coefficients in nonincreasing order, padded with zeros.
Majorization means
\begin{equation}
 \sum_{i=1}^k(\lambda_\psi)_i\leq
 \sum_{i=1}^k(\lambda_\phi)_i\quad(k<d),
 \qquad \sum_i(\lambda_\psi)_i=\sum_i(\lambda_\phi)_i.
\end{equation}
A symmetric function of the spectrum is Schur concave if it reverses
this order. For at most two nonzero components,
$\lambda_x=(\beta_x^2,\alpha_x^2,0,\ldots)$, with
$0\leq\alpha_x\leq\beta_x$, so majorization reduces to
\begin{equation}
 \beta_\psi\leq\beta_\phi,\qquad
 \alpha_\psi\geq\alpha_\phi.
\end{equation}
Equality is permitted, in particular for local-unitary conversions.

\begin{theorem}
For bipartite pure states of Schmidt rank at most two, a deterministic
LOCC conversion $\ket\psi\longrightarrow\ket\phi$ implies
\begin{equation}
 \dhm(\psi)\geq\dhm(\phi),\qquad
 \dhssqm(\psi)\geq\dhssqm(\phi).
\end{equation}
\end{theorem}
\begin{proof}
On the simple branch,
\begin{equation}
 \frac{\mathrm d}{\mathrm d\alpha}\dhssqm
 =\frac83\alpha(1-2\alpha^2)>0
 \quad(\alpha_*\leq\alpha<1/\sqrt2).
\end{equation}
Appendix~\ref{app:derivative} proves strict positivity of the
boundary-branch derivative for $0<\alpha<\alpha_*$. The two values
agree at $\alpha_*$ and the value at zero is zero. Thus
$\dhssqm$ increases with the smaller Schmidt coefficient over the
whole ordered interval. Taking the nonnegative square root preserves
the deterministic ordering. Nielsen's criterion then gives the stated
inequalities, including product outputs.
\end{proof}

\subsection{Entanglement ordering}
Within the class of bipartite pure states of Schmidt rank at most two,
minimum Hilbert--Schmidt distance, concurrence and negativity induce
the same ordering. Indeed, $C=2N=2\alpha\sqrt{1-\alpha^2}$ is
strictly increasing on $0\leq\alpha\leq1/\sqrt2$, as is the minimum
distance by the proof above. Thus, for any two states in this class,
\begin{equation}
\begin{aligned}
 \dhm(\psi)\geq\dhm(\phi)
 &\ \Longleftrightarrow\ C(\psi)\geq C(\phi)\\
 &\ \Longleftrightarrow\ N(\psi)\geq N(\phi).
\end{aligned}
\end{equation}
Equality in any of these comparisons is equivalent to equality of
the nonzero Schmidt spectra. The explicit conversion is given in
Sec.~\ref{sec:concurrence-negativity}; it is linear only on the simple
branch. This ordering equivalence is restricted to the stated
pure-state class.

The endpoint derivative distinction matters. Near the product state,
$\dhssqm=(3/2)\alpha^2+O(\alpha^3)$, so the squared distance has zero
right derivative, whereas $\dhm$ has right derivative $\sqrt{3/2}$.
At the Bell endpoint both active derivatives vanish. The theorem does
not assert an inequality averaged over selective outcomes.

For every mixed-state decomposition $\rho=\sum_i p_i\op{\psi_i}$,
convexity of distance to a convex set gives
\begin{equation}
 \dhm(\rho)\leq\sum_i p_i\dhm(\psi_i).
\end{equation}
The convex roof $\inf_{\{p_i,\psi_i\}}\sum_i p_i\dhm(\psi_i)$
is a different optimization from direct distance to the separable set.
Neither this convexity inequality nor deterministic pure-state
monotonicity alone establishes mixed-state or on-average LOCC
monotonicity of the direct distance.

\begin{figure}[t]
 \includegraphics[width=\linewidth]{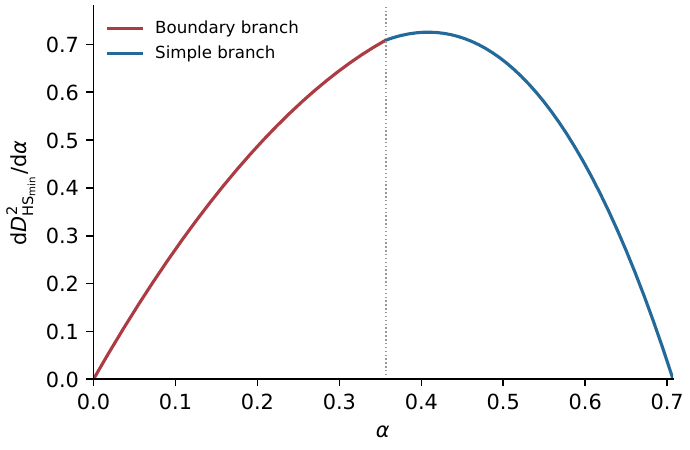}
 \caption{Derivative of the \emph{squared} minimum distance. Red and
 blue show the boundary and simple branches, respectively. The
 dotted line marks $\alpha_*$. The analytical sign proof is given in
 Appendix~\ref{app:derivative}; the plot is a verification.}
 \label{fig:gradient}
\end{figure}

\section{Conclusions and outlook}
We have given a global derivation of the minimum Hilbert--Schmidt
distance for pure states of Schmidt rank at most two. The simple
branch and threshold were established by Witte and Trucks, who also
conjectured the cubic boundary construction. A symmetry reduction and
a trace-preserving saturation argument establish that this construction
minimizes over the full two-qubit PPT set, which equals the separable
set. The compression argument proves that unused local dimensions do
not change even the minimum over the full PPT set. Its unique optimizer
is separable, with an explicit mixture of at most five pure product
states, reducing to three terms on the boundary branch. The closest
state is piecewise in the same two regimes.

The cubic-root bound gives a direct analytical proof of monotonicity
under deterministic pure-state LOCC conversions in this class.
Independent full PPT optimization verifies the expression numerically,
and Rust Gilbert iterations give consistent separable upper bounds.
Selective and mixed-state monotonicity are distinct questions and are
not established here. At higher Schmidt rank PPT optimization remains
a useful lower bound on separable distance. Whether a particular
structured PPT optimizer is separable must be checked separately;
the existence of PPT-entangled states alone does not decide that issue.

\section*{Data availability}
The Python validation and comparison scripts, CSV figure data,
solver-residual records, and raw Gilbert inputs, outputs and trajectories
are available in the manuscript-specific GitHub repository
\texttt{min-HSD-SR-2}, version 0.1.1 \cite{minhsd-code}.
The underlying Rust Gilbert implementation is available in the separate
\texttt{gilbert-opt-rs} repository \cite{gilbert-opt-rs}.
Exact source snapshots, lockfiles, provenance records and
manuscript-specific patches accompany the comparison data, including
the baseline and refined numerical protocols.

\begin{acknowledgments}
This work was supported by the QuantERA II Programme (VERIqTAS project),
which has received funding from the European Union's Horizon 2020
research and innovation programme under Grant Agreement No. 101017733,
and from the Polish National Science Centre
(Grant No. 2021/03/Y/ST2/00175). We also acknowledge funding from the
European Union's Horizon Europe research and innovation programme
under Grant Agreement No. 101080086 (NeQST).

\paragraph*{AI Use Disclosure}: Claude Opus 5.5 and ChatGPT Sol 6.1 assisted with correcting the proof from the previous version of the manuscript, then revising and editing the manuscript, in addition to preparing the accompanying GitHub repository and validation scripts. The author directed the proof revisions and code development, reviewed the resulting arguments and implementations, and verified the analytical results through symbolic checks and independent PPT semidefinite optimization using CVXPY. The author takes responsibility for the final manuscript and accompanying code.

\end{acknowledgments}

\appendix
\section{Symmetry reduction and global optimality}
\label{app:commutator}
\subsection{Commutation from the symmetries of a pure state}
\label{app:schmidt-symmetries}

We first show that the partial transpose of a pure state commutes with the
partial transpose of its closest separable state when the minimization is
performed on its Schmidt support.  This statement holds for any Schmidt
rank~$r$.  Write
\begin{equation}
    \ket\psi=\sum_{i=1}^r\sqrt{\lambda_i}\ket{ii},\qquad
    \sum_i\lambda_i=1,\qquad
    \rho=\op{\psi},
\end{equation}
where $\lambda_i>0$ are the squared Schmidt coefficients.  The Schmidt
decomposition is a choice of local bases, which preserves separability and
Hilbert--Schmidt distance.  For now, both the input and the candidate separable
states act on the corresponding $r\times r$ support; allowing candidates
outside this support is a separate question.

\paragraph{Why the closest state inherits a symmetry.}
Let $\sigma$ minimize $\norm{\rho-\sigma}_2^2$ over the separable states on
this support.  The minimizer exists because this set is compact and is unique
because the set is convex and squared Hilbert--Schmidt distance is strictly
convex.  If a transformation $T$ preserves the separable set and
Hilbert--Schmidt distance and satisfies $T(\rho)=\rho$, then
\begin{equation}
    \norm{\rho-T(\sigma)}_2^2
    =\norm{T(\rho)-T(\sigma)}_2^2
    =\norm{\rho-\sigma}_2^2.
\end{equation}
Thus $T(\sigma)$ is also a minimizer, and uniqueness gives
$T(\sigma)=\sigma$.  Equivalently, averaging a candidate over such symmetries
preserves separability and cannot increase its distance from $\rho$.  We are
therefore using symmetries already possessed by the input, rather than
assuming additional symmetries of the Schmidt coefficients.

\paragraph{The symmetries supplied by Schmidt form.}
First, for arbitrary phases $\theta_i$, define
\begin{equation}
    U_{\boldsymbol\theta}
    =\sum_i e^{i\theta_i}\ket{i}\bra{i}.
\end{equation}
The local unitary $U_{\boldsymbol\theta}\otimes
\overline U_{\boldsymbol\theta}$ leaves $\ket\psi$ unchanged: its two phases
cancel on every $\ket{ii}$.  The CSS must therefore be invariant under all
these local unitaries.  A matrix element $\mel{ij}{\sigma}{kl}$ acquires the
phase $e^{i(\theta_i-\theta_j-\theta_k+\theta_l)}$, so it can survive only
when $i=j$, $k=l$, or when $i=k$, $j=l$.  Consequently,
\begin{equation}
    \sigma=\sum_{i,j}A_{ij}\ket{ii}\bra{jj}
           +\sum_{i\ne j}d_{ij}\ket{ij}\bra{ij}.
\end{equation}
Second, interchange of the two parties on the Schmidt support leaves
$\ket\psi$ unchanged and maps every product state to a product state.  This
unitary symmetry preserves separability and distance, and hence
$d_{ij}=d_{ji}$.  Third, entrywise complex conjugation in the Schmidt basis
leaves $\rho$ unchanged and also preserves separability and distance: a
product vector is mapped to the product of its conjugate local vectors.
Thus $\sigma=\overline\sigma$ and $A_{ij}=A_{ji}\in\mathbb R$.
Complex conjugation is used here as a mathematical symmetry of the
optimization; it need not be a physical quantum channel.  None of these
steps requires equal Schmidt coefficients.

\paragraph{The common eigenbasis.}
Partial transposition maps $\ket{ii}\bra{jj}$ to $\ket{ij}\bra{ji}$.
For every $i<j$, the restrictions to
$\operatorname{span}\{\ket{ij},\ket{ji}\}$ are therefore
\begin{equation}
    \rho^\Gamma\big|_{ij}=
    \begin{pmatrix}
        0&\sqrt{\lambda_i\lambda_j}\\
        \sqrt{\lambda_i\lambda_j}&0
    \end{pmatrix},\qquad
    \sigma^\Gamma\big|_{ij}=
    \begin{pmatrix}
        d_{ij}&A_{ij}\\
        A_{ij}&d_{ij}
    \end{pmatrix}.
\end{equation}
Both blocks are diagonal in the symmetric and antisymmetric vectors
$(\ket{ij}\pm\ket{ji})/\sqrt2$.  The remaining vectors $\ket{ii}$ are
eigenvectors of both operators, with eigenvalues $\lambda_i$ and $A_{ii}$,
respectively.  This gives a common eigenbasis and proves
\begin{equation}
\label{eq:css-pt-commutation}
    [\rho^\Gamma,\sigma^\Gamma]=0.
\end{equation}
The argument also applies to minimization over the PPT set, since that set
is convex, compact and preserved by the same transformations.  This does
not assert that the two minimizers coincide at arbitrary rank.

\subsection{Trace-preserving saturation for rank two}
For rank two, the symmetry-reduced candidate has the form
\begin{equation}
 X=\begin{pmatrix}
 A_{00}&0&0&x\\0&d&0&0\\0&0&d&0\\x&0&0&A_{11}
 \end{pmatrix},\quad
 A=\begin{pmatrix}A_{00}&x\\x&A_{11}\end{pmatrix}.
\end{equation}
The conditions are $A\succeq0$, $d\geq|x|$, and
$A_{00}+A_{11}+2d=1$. Replacing a negative real $x$ by $|x|$
preserves both positivity conditions and reduces its squared error
against the positive input coherence $\alpha\beta$. Hence $x\geq0$
may be imposed without loss.

Suppose $d>x$. Choose $0<h\leq d-x$ and vary
$A\mapsto A+h\id_2$ and $d\mapsto d-h$, keeping $x$ fixed.
The trace is unchanged, $A+h\id_2\succeq0$, and $d-h\geq x$.
The objective before the variation is
\begin{multline}
 F_X=(\alpha^2-A_{00})^2+(\beta^2-A_{11})^2\\
 +2(\alpha\beta-x)^2+2d^2.
\end{multline}
Using normalization, its exact change is
\begin{equation}
 \Delta F_X=-8dh+4h^2<0.
\end{equation}
Thus $d=x$ at the optimum. Setting
$e_1=A_{00}$, $e_2=A_{11}$ and $e_3=2x$ gives exactly
Eq.~\eqref{eq:rank2-state}, with $\sum_i e_i=1$, $e_i\geq0$ and
$4e_1e_2\geq e_3^2$. These constraints describe a convex set, and the
objective is strictly convex in $(e_1,e_2,e_3)$. The reduced minimizer
is therefore the unique full supported PPT minimizer.

For $\alpha\beta>0$, the optimum cannot have $e_3=0$. If both
diagonal entries are positive, the feasible variation
$e_3\mapsto\epsilon$, $e_1,e_2\mapsto e_1-\epsilon/2,e_2-\epsilon/2$
has objective derivative $-2\alpha\beta<0$ at $\epsilon=0$.
The endpoints $(e_1,e_2)=(0,1),(1,0)$ are also excluded by varying
along Eq.~\eqref{eq:boundary-param}; the inward derivatives are
$-4(\alpha^2+\alpha\beta)$ and $-4(\beta^2+\alpha\beta)$,
respectively. Furthermore, PSD forces $e_3=0$ whenever either
$e_1$ or $e_2$ is zero. Hence $e_i>0$ at the entangled-input optimum.

If $4e_1e_2>e_3^2$, the optimum is interior in the trace plane and
must equal the unconstrained minimizer. Below $\alpha_*$ that
minimizer is infeasible. The optimum must therefore satisfy
$4e_1e_2=e_3^2$ there. The boundary calculation in the main text is
thus global. Since PPT is equivalent to separability on $2\times2$,
the supported minimizer is separable.

\subsection{Embedding into larger local Hilbert spaces}
\label{app:embedding}
Let $P=P_A\otimes P_B$ project onto the $2\times2$ Schmidt support,
and let $\tau$ denote the supported optimizer embedded by zero outside
$P$. Put $c_0=\Tr[(\tau-\rho)\tau]$.
The projection condition on the supported separable set gives
\begin{equation}
 c_0\leq\mel{00}{\tau-\rho}{00}=e_1-\alpha^2\leq0.
\end{equation}
The last inequality follows directly from
Eq.~\eqref{eq:simple-coefficients} in the simple region. On the
boundary branch, Lemma~\ref{lem:physical-root} below gives
$e_1=t^2\leq\alpha^2\beta^2/4\leq\alpha^2$. The lemma uses only
the supported boundary minimization, not this embedding argument.

Let $\sigma$ now be any normalized PPT state in the full space and
set $q=\Tr(P\sigma P)\in[0,1]$. If $q>0$, its normalized
compression $\widehat\sigma=P\sigma P/q$ is a PPT state on the
$2\times2$ Schmidt support. In fact,
\begin{equation}
 (P\sigma P)^\Gamma=
 (P_A\otimes\overline{P_B})\sigma^\Gamma
 (P_A\otimes\overline{P_B})\succeq0.
\end{equation}
The two-qubit PPT criterion makes $\widehat\sigma$ separable
\cite{PPT-2}, so supported optimality gives
$\Tr[(\tau-\rho)\widehat\sigma]\geq c_0$.
Since $\tau-\rho$ is supported on $P$, it follows that
\begin{equation}
\begin{aligned}
 \Tr[(\tau-\rho)(\sigma-\tau)]
 &=q\Tr[(\tau-\rho)\widehat\sigma]-c_0\\
 &\geq(q-1)c_0\geq0.
\end{aligned}
\end{equation}
For $q=0$, the left-hand side is $-c_0\geq0$ directly.
This is the projection inequality for the full PPT set. Hence $\tau$
is its unique closest state, by strict convexity of the squared norm.
The explicit decomposition in Sec.~\ref{sec:explicit-decomposition}
also places $\tau$ in the full separable set, proving
Eq.~\eqref{eq:ppt-sep-equality} in arbitrary local dimensions.
The rank-one endpoint is already separable and has zero distance in
every dimension.

\section{The physical cubic root and positivity of the derivative}
\label{app:derivative}
\begin{lemma}
\label{lem:physical-root}
Let $0<\alpha\leq\alpha_*$, $\beta=\sqrt{1-\alpha^2}$ and
$N=\alpha\beta\leq1/3$. The boundary minimizer in
Eq.~\eqref{eq:boundary-param} is the unique root of
Eq.~\eqref{eq:boundary-cubic} in $0<t<1/2$, and satisfies
\begin{equation}
\label{eq:root-bound}
 0<t\leq N/2\leq1/6.
\end{equation}
\end{lemma}
\begin{proof}
Write
\begin{equation}
 p(t)=6t^3-9t^2+(4+2N)t-\alpha^2-N,
 \qquad \partial_tF=4p(t).
\end{equation}
We have $p(0)<0$ and $p(1/2)=(\beta^2-\alpha^2)/2>0$.
Moreover, $p''(t)=36t-18\leq0$ on $[0,1/2]$.
Concavity and these endpoint signs imply that $p$ has exactly one
zero there, crossed with positive slope. Consequently $F$ decreases
up to that root and increases afterwards on this half-interval.
For $t>1/2$,
\begin{equation}
 F(\alpha,t)-F(\alpha,1-t)
 =2(\beta^2-\alpha^2)(2t-1)>0,
\end{equation}
so no point on the other half-interval improves the minimum.

Finally, direct substitution gives
\begin{equation}
\label{eq:root-test}
 p(N/2)=\frac N4(1-3N)(4-N)+(2N^2-\alpha^2)\geq0.
\end{equation}
Both terms are nonnegative: $N\leq1/3$ and
$2N^2-\alpha^2=\alpha^2(2\beta^2-1)\geq0$.
The unique root must therefore lie between $0$ and $N/2$.
\end{proof}

For clarity, the active branches are
\begin{equation}
\label{eq:min-hsd-def-app}
 \dhssqm=\begin{cases}
 F(\alpha,t(\alpha)),&0<\alpha<\alpha_*,\\
 \frac43\alpha^2(1-\alpha^2),&\alpha_*\leq\alpha\leq1/\sqrt2,
 \end{cases}
\end{equation}
with value zero at the product endpoint. The coefficient on the lower
branch is equivalently
\begin{equation}
\label{eq:e2min-app}
 e_2=(1-t)^2,
\end{equation}
whose radical expression is Eq.~\eqref{eq:e2min}.

At the physical root, $\partial_tF=0$ and $p'(t)>0$, so the implicit
function theorem gives a smooth root on the open branch. Differentiating
the minimized objective therefore eliminates $t'(\alpha)$:
\begin{align}
\label{eq:envelope-derivative}
 \frac{\mathrm d\dhssqm}{\mathrm d\alpha}
 &=\partial_\alpha F(\alpha,t)\nonumber\\
 &=\frac4\beta\left[N(1-2t)
 -(\beta^2-\alpha^2)t(1-t)\right].
\end{align}
Put $\delta=\beta^2-\alpha^2\in[0,1]$. Using
Eq.~\eqref{eq:root-bound}, the complete bracket satisfies
\begin{align}
 N(1-2t)-\delta t(1-t)
 &\geq N(1-2t)-t(1-t)\nonumber\\
 &\geq N(1-2t)-\frac N2(1-t)\nonumber\\
 &=\frac N2(1-3t)>0.
\label{eq:derivative-positive}
\end{align}
This proves the required derivative sign without comparing separately
positive and negative terms from differentiated radicals.

At $\alpha_*$ the coefficients approach
Eq.~\eqref{eq:simple-coefficients}, and both values equal $4/27$.
Equation~\eqref{eq:envelope-derivative} also agrees there with the
simple-branch derivative $8\alpha(1-2\alpha^2)/3$: at the junction
$N=1/3$, $t=\alpha_*^2$ and $t(1-t)=1/9$, so both derivatives equal
$8(1-2\alpha_*^2)/(9\beta)$.
Near zero, Eq.~\eqref{eq:boundary-cubic} gives
$t=\alpha/4+O(\alpha^2)$; substituting in
Eq.~\eqref{eq:boundary-objective} yields
$\dhssqm=(3/2)\alpha^2+O(\alpha^3)$.
The endpoint and junction behavior therefore agree with the
piecewise ordering used in the main proof.

\bibliography{apssamp}
\end{document}